\documentstyle[12pt,aps]{revtex}
\newcommand{\bold}[1]{\mbox{\boldmath $#1$}}
\tighten
\begin{document}
\draft

\title{\flushleft Geometric factors in the 
Bohr--Rosenfeld analysis of the measurability of the electromagnetic field}
\author{\flushleft V Hnizdo\footnote{Present address: National Institute
for Occupational Safety and Health, 1095 Willowdale Road, Morgantown, 
WV 26505, USA}}
\address{\flushleft
Department of Physics, Schonland Research Centre for
Nuclear Sciences, and Centre for Nonlinear Studies,
University of the Witwatersrand, Johannesburg, 2050 South Africa}
\maketitle
\begin{abstract}
The geometric factors in the field commutators and spring constants of 
the  measurement devices 
in the famous analysis of the measurability of the electromagnetic 
field by Bohr and Rosenfeld are calculated using a Fourier--Bessel
method for the evaluation of folding integrals, which 
enables one to obtain the general geometric factors  
as a Fourier--Bessel series. 
When the space regions over which the factors are defined are spherical, 
the Fourier--Bessel series terms are given by elementary functions, and
using the standard Fourier-integral method of calculating folding integrals,
the geometric factors can be evaluated in terms of manageable closed-form 
expressions. 
\end{abstract}
\noindent
{\it J. Phys. A: Math. Gen.} {\bf 32} (1999) 2427--2445
\begin{flushleft}
\bf 1. Introduction
\end{flushleft}
The fundamental importance of the famous paper of Bohr and Rosenfeld \cite{BR}
on the measurability of the electromagnetic field is acknowledged by most
physicists but, curiously, the paper seems to have been read by only a
few at the time of its appearance in the early 1930's,\footnote{
A. Pais says in his book on N. Bohr \cite{AP}: ``From decades of involvement
with quantum field theory I can testify that nevertheless it has been read
by  very very few of the {\it aficionados}. The main reason is, I think,
that even by Bohr's
standards this paper is very difficult to penetrate. It takes inordinate
care and patience to follow Bohr's often quite complex gyrations with test
bodies{...} As a friend of Bohr and mine once said to me: `It is a very good
paper that one does not have to read. You just have to know it exists.'
Nevertheless men like Pauli and Heitler did read it with great care."}
and undoubtedly by still less in the more recent times.\footnote{
A contributing factor to that must have been that the paper had not been
available in an English translation for a long time.}
The Bohr--Rosenfeld (BR) paper arose in a response to
Landau and Peierls \cite{LP}, who argued that, in principle,
the electromagnetic field is not measurable in a domain where quantum and
relativistic effects are important. In their paper, BR have refuted this
claim by  showing that in quantum electrodynamics,
just as in nonrelativistic quantum mechanics, there is a complete harmony
between the theoretical formalism and the physical possibilities of
measurement.
The essential aspects of the BR analysis that enabled them to reach
this conclusion are the realization that only field quantities averaged
over finite space-time regions are physically meaningful,
and the employment of essentially classical test bodies of finite
size that can have arbitrarily large charge and mass, which removed 
immediately the limit on the field measurability
due to the radiation-reaction force on the point test charges employed by 
Landau and Peierls. 
BR have shown that the electromagnetic-field 
effects of such finite-size test bodies can be minimized to exactly
the extent demanded by the formalism's commutation relations by using
classical spring and lever mechanisms connecting the
test bodies to the frame of reference and to each other, together
with the deployment of neutralizing bodies that occupy the same space
regions as the test bodies but which are charged oppositely and remain 
attached 
rigidly to the reference frame during the duration of a measurement.
The harmony between the possibilities of definition afforded by the
quantum-electrodynamic formalism and the possibilities of measurement
could be attained only by the masterful exploitation by BR of all the
opportunities offered to measurement by classical physics while
at the same time paying due regard to the limitations imposed on the latter
by quantum  mechanics.\footnote{In the course of their analysis, BR had
to examine the problem of measurement of the basic mechanical quantities   
of position, momentum and energy in more detail than it had been done in the
previous writings of Bohr and Heisenberg, and established the possibility
of repeatable momentum and energy measurements that may be of arbitrarily
short duration, which were rediscovered some 30 years later by Aharonov 
and Bohm \cite{AB}.}

Several illuminating commentaries on the BR analysis have been
written \cite{AP,Rosen,Peierls,Miller}, some of them by authors who were 
close to the original controversy. While the BR field measurement philosophy 
has not been accepted unreservedly by all the writers \cite{Peierls}, 
the technical correctness of the BR analysis has not been disputed.
Very recently, however, an analysis \cite{CP}  of the BR procedure for the 
measurement 
of a single space-time-averaged component of the electromagnetic 
field has drawn a conclusion that no compensating spring mechanism
is needed in order to measure the averaged field component to arbitrary 
accuracy when no neutralizing body is employed. 
This work is commented on critically
elsewhere \cite{Com}, using the calculational methods developed in the 
present paper.

The field commutation relations that BR use as the starting point 
of their analysis, and the spring constants of the mechanisms employed 
in their measurement procedures, are defined in terms of geometric factors
that are double averages over two
finite four-dimensional space-time regions. As this amounts formally to an
eight-dimensional integration, the calculation of the value of 
a field commutator for a finite space-time region, 
or of a BR spring constant, is not simple, even though 
the dimensionality of the integration is reduced, albeit not in a 
straightforward manner, by the presence of a delta function in the integrand.
To the present author's knowledge, no calculations of these quantities 
have been reported in the literature yet, apart from those for coinciding
spherical space-time regions in the comment \cite{Com} on 
reference \cite{CP}.
Clearly, a well-controlled algorithm for the evaluation
of the field commutators for finite space-time regions and the BR spring 
constants is desirable---not least 
because, as we shall see, these quantities amount essentially to the field 
effects of extended charged bodies, a quantitative assessment of which
may be needed in special experimental situations.\footnote{
For example,
direct detection of gravitational waves would require repeated  measurements 
of very high accuracy on a single object in a regime where quantum effects 
are important. In these so-called ``quantum nondemolition" measurements, 
experimental precision is pushed to the limits set by the principles of 
quantum mechanics and quantum electrodynamics; under such 
or similar circumstances, the measurement procedures and results of the 
BR analysis may well be of practical relevance \cite{QNM}.}

In the
present paper, the Fourier--Bessel method developed for an efficient and
accurate evaluation of multiple folding integrals \cite{VH} is adapted 
for the calculation of general BR geometric factors. However,
the standard Fourier-integral method for calculating folding integrals
will turn out to be more advantageous in the special case of spherical 
space regions, as it will enable us to evaluate the BR geometric 
factors with spherical space regions in closed form. 

In section 2 of this paper, 
the BR field commutators are introduced, and the Fourier--Bessel method for 
calculating the general geometric factors in terms of which the field 
commutators and the BR spring constants are defined is developed. 
In section 3,  
geometric factors with spherical space regions are considered; these are 
first calculated using Fourier--Bessel expansions, and then evaluated  
in closed form. In the last section, some calculational results are 
presented and discussed, and concluding remarks are made.

\begin{flushleft}
\bf 2. Fourier--Bessel expansions of the BR geometric factors
\end{flushleft}

The starting point in the BR analysis is the set of electromagnetic-field
commutation relations for the operators of field averages over finite 
space-time regions instead of for those of the field values at space-time
points:
\begin{eqnarray}
&&[\bar{\cal E}^{\rm(I)}_x,\bar{\cal E}^{\rm(II)}_x]
=[\bar{\cal H}^{\rm(I)}_x,\bar{\cal H}^{\rm(II)}_x]
={\rm i}\hbar(\bar{A}^{\rm(I,II)}_{xx}-\bar{A}^{\rm(II,I)}_{xx}), \\ 
&&[\bar{\cal E}^{\rm(I)}_x,\bar{\cal E}^{\rm(II)}_y]
=[\bar{\cal H}^{\rm(I)}_x,\bar{\cal H}^{\rm(II)}_y]
={\rm i}\hbar(\bar{A}^{\rm(I,II)}_{xy}-\bar{A}^{\rm(II,I)}_{xy}), \\ 
&&[\bar{\cal E}^{\rm(I)}_x,\bar{\cal H}^{\rm(II)}_x]=0, \\
&&[\bar{\cal E}^{\rm(I)}_x,\bar{\cal H}^{\rm(II)}_y]
=-[\bar{\cal H}^{\rm(I)}_x,\bar{\cal E}^{\rm(II)}_y]
={\rm i}\hbar(\bar{B}^{\rm(I,II)}_{xy}-\bar{B}^{\rm(II,I)}_{xy}).
\end{eqnarray}
Here, $\bar{\cal E}^{\rm(I)}_x, \bar{\cal H}^{\rm(I)}_x$, etc.\ are
the electric and magnetic field components averaged over a space-time
region I of volume $V_{\rm I}$ and duration $T_{\rm I}$, as, for example, 
\begin{equation}
\bar{\cal E}^{\rm(I)}_x=\frac{1}{V_{\rm I}T_{\rm I}}\int_{T_{\rm I}}{\rm d}
t_1 \int_{V_{\rm I}}{\rm d}v_1\,{\cal E}_x(x_1,y_1,z_1,t_1),
\end{equation}
and the right-hand-side quantities are purely {\it geometric factors} 
defined in terms of double averages over space-time regions I and II: 
\begin{eqnarray}
\bar{A}^{(\rm I,II)}_{xx}&=& -\frac{1}{\cal V_{\rm I,II}}
\int_{T_{\rm I}}\!{\rm d}t_1\int_{T_{\rm II}}\!{\rm d}t_2
\int_{V_{\rm I}}\!{\rm d}v_1\int_{V_{\rm II}}\!{\rm d}v_2
\left(\frac{\partial^2}{\partial x_1\partial x_2}
-\frac{1}{c^2}\frac{\partial^2}{\partial t_1\partial t_2}\right)
\left[\frac{1}{r}\delta\left(t_2-t_1-\frac{r}{c}\right)\right],\\
\bar{A}^{(\rm I,II)}_{xy}&=& -\frac{1}{\cal V_{\rm I,II}}
\int_{T_{\rm I}}\!{\rm d}t_1\int_{T_{\rm II}}\!{\rm d}t_2
\int_{V_{\rm I}}\!{\rm d}v_1\int_{V_{\rm II}}\!{\rm d}v_2\,
\frac{\partial^2}{\partial x_1\partial y_2}
\left[\frac{1}{r}\delta\left(t_2-t_1-\frac{r}{c}\right)\right],\\
\bar{B}^{(\rm I,II)}_{xy}&=&-\frac{1}{\cal V_{\rm I,II}}
\int_{T_{\rm I}}\!{\rm d}t_1\int_{T_{\rm II}}\!{\rm d}t_2
\int_{V_{\rm I}}\!{\rm d}v_1\int_{V_{\rm II}}\!{\rm d}v_2\,
\frac{1}{c}\frac{\partial^2}{\partial t_1\partial z_2}
\left[\frac{1}{r}\delta\left(t_2-t_1-\frac{r}{c}\right)\right],
\end{eqnarray}
where ${\cal V}_{\rm I,II}=V_{\rm I}V_{\rm II}T_{\rm I}T_{\rm II} $
and $r$ is the distance between a space point in the region I and a space
point in the region II. The remaining commutation relations are obtained
from (1)--(4) and (6)--(8) by cyclic permutations.

A BR geometric factor, say
$\bar{C}^{(\rm I,II)}_U$,  can be written as
\begin{equation}
\bar{C}^{(\rm I,II)}_U =\frac{1}{\Delta t_1 \Delta t_2}
\int_0^{\Delta t_1}{\rm d}t_1\int_{T}^{T+\Delta t_2}{\rm d}t_2\, 
\int\rho_1(\bold{r}_1){\rm d}\bold{r}_1\int\rho_2(\bold{r}_2){\rm d}
\bold{r}_2\, U(t,\bold{r}). 
\end{equation}
Here, the time intervals associated with the space-time regions I and II 
are specified, without loss of generality, 
as $(0,\Delta t_1)$ and $(T,T+\Delta t_2)$,\footnote{ 
BR use the symbols $T_{\rm I}$ and $T_{\rm II}$ for these time intervals, 
and the symbol $\Delta t$ for the durations  
of the momentum measurements at the beginning and end of 
a field-measurement time interval.}
respectively, while the space regions are given by the means of uniform 
density distributions $\rho_1(\bold{r}_1)$ and $\rho_2(\bold{r}_2)$ that 
vanish outside the space regions I and II, respectively, 
and are each normalized to unit volume; 
the coordinates $\bold{r}_1$ and $\bold{r}_2$ now refer to origins located
conveniently inside the regions I and II so that the displacement $\bold{r}$ 
of a space point of region II from a space point of region I is given as 
$\bold{r}=\bold{R}+\bold{r}_2-\bold{r}_1$, with $\bold{R}$ the displacement 
of the origin of region II from that of region I.
The function $U(t,\bold{r})$ in equation (9) is the integrand of 
the multiple 
integral defining the geometric factor; for the geometric factors (6)--(8),
it takes the following forms, respectively:
\begin{eqnarray}
U_{A_{xx}}(t,\bold{r})
&=&-\left(\frac{\partial^2}{\partial x_1\partial x_2}
-\frac{\partial^2}{\partial t_1\partial t_2}\right)
\frac{\delta(t-r)}{r}, \\  
U_{A_{xy}}(t,\bold{r})
&=&-\frac{\partial^2}{\partial x_1\partial y_2}
\frac{\delta(t-r)}{r}, \\  
U_{B_{xy}}(t,\bold{r})
&=&-\frac{\partial^2}{\partial t_1\partial z_2}
\frac{\delta(t-r)}{r},   
\end{eqnarray}
where units such that the speed of light $c=1$ are now used, 
and $t=t_2-t_1$.
The BR geometric factor (9) is cast in the form of a time average
of a double-folding integral whose integrand involves functions that
all have finite space extension, as the densities $\rho_k(\bold{r}_k)$
represent finite regions of space, and the radial range of the 
function $U(t,\bold{r})$ is given by $r=|\bold{r}|= t$ as it contains 
the delta function $\delta(t-r)$ and its derivatives.
As such, the BR geometric factors are particularly well suited to 
evaluation by the means of Fourier--Bessel expansions \cite{VH}.

To this end, the multipoles $U_{lm}(t,r)$ of the nonspherical function 
$U(t,\bold{r})$, defined by a multipole expansion
\begin{equation}
U(t,\bold{r})=\sum_{lm}U_{lm}(t,r){\rm i}^lY_{lm}(\hat{\bold{r}}),
\end{equation}
are expanded as Fourier--Bessel series in 
spherical Bessel functions $j_l(q_n^{(l)}r)$ in a range $0\le r<r_{\rm ex}$: 
\begin{equation}
U_{lm}(t,r)=\sum_{n=1}^{\infty}c^{(lm)}_{U\,n}(t)j_l(q_n^{(l)}r), 
\end{equation}
where $q_n^{(l)}r_{\rm ex}$ are the positive roots of $j_l(x)$.
The coefficients $c_{U\,n}^{(lm)}(t)$ in (14) 
are given  in terms of the multipoles $U_{lm}(t,r)$ as
\begin{equation}
c_{U\,n}^{(lm)}(t)=
\frac{1}{w_n^{(l)}}\int_0^{r_{\rm ex}}U_{lm}(t,r)j_l(q_n^{(l)}r)r^2
\,{\rm d}r,
\end{equation}
where
\begin{equation}
w_n^{(l)}=\frac{r_{\rm ex}}{2}[r_{\rm ex}j_l'(q^{(l)}_n r_{\rm ex})]^2.
\end{equation}
The multipole expansion (13) can thus be written for $|\bold{r}|< r_{\rm ex}$
as
\begin{equation}
U(t,\bold{r})
=\frac{1}{4\pi}\sum_{lm}\sum_{n=1}^{\infty}c_{U\,n}^{(lm)}(t)
\int\exp({\rm i}\bold{q}_n^{(l)}{\bf\cdot}\bold{r})
Y_{lm}(\hat{\bold{q}}_n^{(l)})\,{\rm d}\hat{\bold{q}}_n^{(l)}, 
\end{equation}
where  $\bold{q}_n^{(l)}$ are vectors with polar angles 
$\hat{\bold{q}}_n^{(l)}$ and discrete moduli 
$|\bold{q}_n^{(l)}|=q_n^{(l)}$, and where the identity
\begin{equation}
j_l(qr){\rm i}^lY_{lm}(\hat{\bold{r}})
=\frac{1}{4\pi}
\int{\rm exp}({\rm i}\bold{q}{\bf\cdot}\bold{r})Y_{lm}(\hat{\bold{q}})\,
{\rm d}\hat{\bold{q}}
\end{equation}
is employed.

In the double-folding integral of equation (9), the value of the function 
$U(t,\bold{r})$ is required only when 
\begin{equation}
|\bold{r}|=|\bold{R}+\bold{r}_2-\bold{r}_1|\le r_{\rm max}=R+R_1+R_2,
\end{equation}
where $R$ is the separation of the centres of the two densities, and
$R_1$ and $R_2$ are the radii beyond which the uniform
densities $\rho_1(\bold{r}_1)$ and $\rho_2(\bold{r}_2)$, respectively,  
vanish---this is simply because the product 
$\rho_1(\bold{r}_1)\rho_2(\bold{r}_2)$ in the integrand of (9) is bound to 
be zero when $|\bold{r}| > r_{\rm max}$. 
Thus, the validity of the Fourier--Bessel expansion (17) 
in the double-folding 
integral of equation (9) will be guaranteed when the expansion radius 
$r_{\rm ex}\ge r_{\rm max}$.   
Substituting then the expansion (17) 
with $\bold{r}=\bold{R}+\bold{r}_2-\bold{r}_1$ and an expansion radius
$r_{\rm ex}\ge r_{\rm max}$ in equation (9), one obtains 
a Fourier--Bessel expansion for the BR geometric factor
$\bar{C}^{(\rm I,II)}_U$:
\begin{equation}
\bar{C}^{(\rm I,II)}_U 
=\frac{1}{4\pi}\sum_{lm}\sum_{n=1}^\infty\bar{c}^{(lm)}_{U\,n} \int
\tilde{\rho}_1(-\bold{q}^{(l)}_n)\tilde{\rho}_2
(\bold{q}^{(l)}_n){\rm exp}({\rm i}\bold{q}^{(l)}_n
{\bf\cdot}\bold{R}) Y_{lm}(\hat{\bold{q}}^{(l)}_n )\,
{\rm d}\hat{\bold{q}}^{(l)}_n.
\label{FBV}
\end{equation}
Here, 
\begin{equation}
\bar{c}_{U\,n}^{(lm)}=
\frac{1}{\Delta t_1\Delta t_2}\int_0^{\Delta t_1}{\rm d}t_1
\int_T^{T+\Delta t_2}{\rm d}t_2\,c_{U\,n}^{(lm)}(t),
\end{equation}
are the time averages of the coefficients (15),
and $\tilde{\rho}_1(-\bold{q}^{(l)}_n)$ and 
$\tilde{\rho}_2(\bold{q}^{(l)}_n)$ are the Fourier transforms of 
the densities $\rho_k(\bold{r}_k)$,
\begin{equation}
\tilde{\rho}_k(\bold{q})=
\int\rho_k(\bold{r}_k)\,{\rm exp}({\rm i}\bold{q}{\bf\cdot}\bold{r}_k)
\,{\rm d}\bold{r}_k,
\end{equation}
evaluated at the points $\bold{q}=-\bold{q}^{(l)}_n$ and $\bold{q}^{(l)}_n$, 
respectively.

The Fourier transforms $\tilde{\rho}_k(\bold{q})$ can be expanded in 
multipoles also,
\begin{equation}
\tilde{\rho}_k(\bold{q})=\sum_{l_km_k}\tilde{\rho}_{l_km_k}^{(k)}(q)
{\rm i}^{l_k}Y_{l_km_k}(\hat{\bold{q}}),
\end{equation}
which are given in terms of the similarly defined multipoles 
$\rho_{l_km_k}^{(k)}(r_k)$
of the densities $\rho_k(\bold{r}_k)$ themselves by
\begin{equation}
\tilde{\rho}_{l_km_k}^{(k)}(q)=4\pi{\rm i}^{l_k}\int_0^{\infty}
\rho_{l_km_k}^{(k)}(r_k)j_{l_k}(qr_k)r_k^2{\rm d}r_k.
\end{equation}
When the multipole expansions (23) are substituted in (20), the products 
of spherical harmonics are expanded in terms of single spherical harmonics,
and the identity (18) is used again,
the Fourier--Bessel expansion of the BR geometric factor  
$\bar{C}^{(\rm I,II)}_U$ can be written finally as\footnote{
Cf.\ equation (21) 
of reference \cite{VH}, which is reconciled with our equation (25) on noting 
that our displacement $\bold{r}=\bold{R}+\bold{r}_2-\bold{r}_1$
is defined there as $\bold{r}=\bold{R}+\bold{r}_1-\bold{r}_2$, and that 
$[\tilde{\rho}_{lm}(q)]^*=(-1)^m \tilde{\rho}_{l-m}(q)$
for a real density $\rho(\bold{r})$.} 
\begin{eqnarray}
&&\bar{C}^{(\rm I,II)}_U=\frac{1}{4\pi}\sum_{lm}
\sum_{\scriptstyle l_1m_1\atop \scriptstyle l_2m_2}
\sum_{n=1}^\infty \bar{c}^{(lm)}_{U\,n}(-1)^m\hat{l}
\hat{l}_1\hat{l}_2{\rm i}^{-l_1}
\tilde{\rho}^{(1)}_{l_1m_1}\!\big(q^{(l)}_n\big){\rm i}^{l_2}
\tilde{\rho}^{(2)}_{l_2m_2}\!\big(q^{(l)}_n\big) 
\nonumber \\ 
&& \times\sum_{\lambda'\lambda}\hat{\lambda}'^2\hat{\lambda}
\Bigg(
\!\begin{array}{ccc}l_1&l_2&\lambda'\\m_1&m_2&-\mu'\end{array}\!\Bigg)
\Bigg(
\!\begin{array}{ccc}l_1&l_2&\lambda'\\0 & 0 & 0  \end{array}\!\Bigg)
\Bigg(
\!\begin{array}{ccc}\lambda'&l&\lambda\\ \mu'&m &-\mu\end{array}\!\Bigg)
\Bigg(
\!\begin{array}{ccc}\lambda'&l&\lambda\\0 & 0 & 0  \end{array}\!\Bigg)
j_{\lambda}\big(q^{(l)}_nR\big){\rm i}^{\lambda}
Y_{\lambda\mu}\big(\hat{\bold{R}}\big),
\label{FFBY}
\end{eqnarray}
where $\hat{l}=(2l+1)^{1/2}$ etc., and the large parenthesis denote
3-$j$ coefficients; $\mu'=m_1+m_2$ and $\mu=m+m_1+m_2$.
In this way, when the multipole expansions of the function $U(t,\bold{r})$ 
and the densities $\rho_k(\bold{r}_k)$ are given, 
the evaluation of the eight-dimensional integral (9) is reduced
to the evaluation of the one-dimensional integrals (15) for 
$c_{U\,n}^{(lm)}(t)$
and (24) for $\tilde{\rho}_{l_km_k}^{(k)}(q_n^{(l)})$, 
of the two-dimensional integrals (21) for the time averages 
$\bar{c}_{U\,n}^{(lm)}$, and
of the Fourier--Bessel expansion (25), where, in principle, only the number 
of terms in the expansion controls the degree of approximation to the exact 
value of the BR geometric factor $\bar{C}^{(\rm I,II)}_U$. 

We now turn to the calculation of the Fourier--Bessel coefficients (15)
and their time averages (21).
It turns out that with the functional forms (10)--(12) taken by 
$U(t,\bold{r})$, 
these quantities can be evaluated in terms of elementary functions.
Let us first determine the multipole expansion of the function 
$U_{A_{xx}}(t,\bold{r})$, given by equation (10).
First, we regularize the space derivative part of $U_{A_{xx}}(t,\bold{r})$ by
\begin{equation}
\frac{\partial^2}{\partial x_1\partial x_2}\frac{\delta(t-r)}{r}
=\lim_{\epsilon\rightarrow 0}
\frac{\partial^2}{\partial x_1\partial x_2}\frac{\delta(t-r)}{r+\epsilon},
\end{equation}
where the limit $\epsilon\rightarrow 0$ is understood to be taken only
after a double integration.
This yields
\begin{eqnarray}
\frac{\partial^2}{\partial x_1\partial x_2}\,\frac{\delta(t-r)}{r}
=&-&\lim_{\epsilon\rightarrow 0}\left\{\left[\frac{\delta''(t-r)}{r+\epsilon}
+\frac{3r+\epsilon}{r(r+\epsilon)^2}\delta'(t-r) 
+\frac{3r+\epsilon}{r(r+\epsilon)^3}\delta(t-r)
\right]\frac{(x_2-x_1)^2}{r^2}\right.\nonumber \\  
&-&\left.\frac{\delta'(t-r)}{r(r+\epsilon)}-\frac{\delta(t-r)}
{r(r+\epsilon)^2}\right\},
\end{eqnarray}
while the time derivatives give 
\begin{equation}
\frac{\partial^2}{\partial t_1\partial t_2}\,\frac{\delta(t-r)}{r}
=-\frac{\delta''(t-r)}{r}.
\end{equation}
Now
\begin{equation}
\frac{1}{r^2}(x_2-x_1)^2=
\case{2\pi}{3}[Y_{1-1}(\hat{\bold{r}})-Y_{11}(\hat{\bold{r}})]^2
=\case{1}{3}+\sqrt{\case{2\pi}{15}}Y_{2-2}(\hat{\bold{r}})
-\case{1}{3} \sqrt{\case{4\pi}{5}}Y_{20}(\hat{\bold{r}})
+\sqrt{\case{2\pi}{15}}Y_{22}(\hat{\bold{r}}).
\end{equation}
Using this in (27), then substituting (27) and (28) 
in (10) and taking the limit $\epsilon\rightarrow 0$ everywhere 
except in the $\delta(t-r)$ term of the monopole component, 
we obtain the function $U_{A_{xx}}(t,\bold{r})$ as a multipole sum
\begin{eqnarray}
&&U_{A_{xx}}(t,\bold{r})=-\frac{2}{3}\sqrt{4\pi}\left[
\frac{\delta''(t-r)}{r}+\lim_{\epsilon\rightarrow 0}
\frac{\epsilon}{(r+\epsilon)^3}\frac{\delta(t-r)}{r}\right]
{\rm i}^0Y_{00}(\hat{\bold{r}})\nonumber \\
&&-\sqrt{\frac{2\pi}{15}}
\left[\frac{\delta''(t-r)}{r}+3\frac{\delta'(t-r)}{r^2} 
+3\frac{\delta(t-r)}{r^3}\right]  
{\rm i}^2\left[Y_{2-2}(\hat{\bold{r}})
-\sqrt{\case{2}{3}}Y_{20}(\hat{\bold{r}}) +Y_{22}(\hat{\bold{r}}) \right].    
\end{eqnarray}
The limit $\epsilon\rightarrow 0$ was performed in the multipole
expansion of $U_{A_{xx}}(t,\bold{r})$ everywhere except in the
$\delta(t-r)$ term of the monopole component because it can be seen easily 
that it is only with this term that the regularization (26) can contribute 
to the geometric factor $A_{xx}^{\rm(I,II)}$.
No regularization is required in the functions (11) and (12), which turn 
out not to contain any monopole components:
\begin{eqnarray} 
U_{A_{xy}}(t,\bold{r})&=& 
-{\rm i}\sqrt{\frac{2\pi}{15}}\left[\frac{\delta''(t-r)}{r}
+3\frac{\delta'(t-r)}{r^2} 
+3\frac{\delta(t-r)}{r^3}\right]{\rm i}^2[Y_{2-2}(\hat{\bold{r}})
-Y_{22}(\hat{\bold{r}})],
\\   
U_{B_{xy}}(t,\bold{r})&=&
{\rm i}\sqrt{\frac{4\pi}{3}}\left[\frac{\delta''(t-r)}{r}
+\frac{\delta'(t-r)}{r^2}\right]{\rm i}Y_{10}(\hat{\bold{r}}).
\end{eqnarray}

The multipoles $U_{lm}(r,t)$ of the various forms
of $U(t,\bold{r})$ are found easily from equations (30)--(32), and 
the integrals required for the  
monopole $(l=0)$, dipole $(l=1)$ and quadrupole $(l=2)$ 
Fourier--Bessel coefficients (15) are evaluated as follows:
\begin{eqnarray} 
&&\int_0^{r_{\rm ex}}\left[\frac{\delta''(t-r)}{r}
+\lim_{\epsilon\rightarrow 0}\frac{\epsilon}{(r+\epsilon)^3}
\frac{\delta(t-r)}{r}\right] j_0(qr)r^2{\rm d}r\nonumber \\
&&\;\;\;\;\;\;=-q\sin(qt)\Theta(t)\Theta(r_{\rm ex}-t)+
\Delta^{(0)}(q,r_{\rm ex},t)+\frac{1}{q}\lim_{\epsilon\rightarrow 0}
\frac{\epsilon\sin(qt)}{(t+\epsilon)^3}
\Theta(t)\Theta(r_{\rm ex}-t), \\
&&\int_0^{r_{\rm ex}}\left[\frac{\delta''(t-r)}{r}
+\frac{\delta'(t-r)}{r^2}\right] j_1(qr)r^2{\rm d}r=
q\cos(qt)\Theta(t)\Theta(r_{\rm ex}-t)+\Delta^{(1)}(q,r_{\rm ex},t),\\
&&\int_0^{r_{\rm ex}}\left[\frac{\delta''(t-r)}{r}+3\frac{\delta'(t-r)}{r^2} 
+3\frac{\delta(t-r)}{r^3}\right]j_2(qr)r^2{\rm d}r  
=q\sin(qt)\Theta(t)\Theta(r_{\rm ex}-t)\nonumber \\
&&\;\;\;\;\;\;\;\;+\Delta^{(2)}(q,r_{\rm ex},t).
\end{eqnarray}
Here,
\begin{eqnarray}
&&\Delta^{(0)}(q,r_{\rm ex},t)
=\delta(t)- \cos(qr_{\rm ex})\delta(t-r_{\rm ex})
-r_{\rm ex} j_0(qr_{\rm ex}) \delta'(t-r_{\rm ex}), \\
&&\Delta^{(1)}(q,r_{\rm ex},t)=-\sin(qr_{\rm ex})\delta(t-r_{\rm ex})
-r_{\rm ex}j_1(qr_{\rm ex})\delta'(t-r_{\rm ex}), \\
&&\Delta^{(2)}(q,r_{\rm ex},t)=[\cos(qr_{\rm ex})-j_0(qr_{\rm ex})
-j_2(qr_{\rm ex})]\delta(t-r_{\rm ex})
-r_{\rm ex}j_2(qr_{\rm ex})\delta'(t-r_{\rm ex}).
\end{eqnarray}
These results are obtained following the rules that govern the use of the 
delta function and its derivatives:\footnote{
As a multidimensional integration with finite
integration limits is involved here, care has to be taken
to use the derivatives of the delta function properly.} 
\begin{eqnarray}
&&\int_{x_1}^{x_2}f(x)\delta(x-x_0)\,{\rm d}x
=f(x_0)\Theta(x_0-x_1)\Theta(x_2-x_0),\\
&&\int_{x_1}^{x_2}f(x)\delta'(x-x_0)\,{\rm d}x
=-f'(x_0)\Theta(x_0-x_1)\Theta(x_2-x_0)\nonumber \\
&&\;\;\;\;+f(x_2)\delta(x_2-x_0)-f(x_1)\delta(x_1-x_0),\\
&&\int_{x_1}^{x_2}f(x)\delta''(x-x_0)\,{\rm d}x
=f''(x_0)\Theta(x_0-x_1)\Theta(x_2-x_0)\nonumber \\
&&\;\;\;\;-f'(x_2)\delta(x_2-x_0) +f'(x_1)\delta(x_1-x_0)
+f(x_2)\delta'(x_2-x_0)-f(x_1)\delta'(x_1-x_0),
\end{eqnarray}
where $\Theta(t)$ is the Heaviside step function: $\Theta(t)=1$ for $t>0$, 
and $\Theta(t)=0$ for $t< 0$; and utilizing the fact that 
$\lim_{x\rightarrow 0}j_l(x)=\lim_{x\rightarrow 0}d[xj_l(x)]/dx=\delta_{0l}$. 

Using equations (30)--(35), the time averaged Fourier--Bessel 
coefficients (21) needed in (25) for 
the BR geometric factor $\bar{A}_{xx}^{\rm(I,II)}$ are thus given by 
\begin{eqnarray}
\bar{c}^{(00)}_{A_{xx}\,n}&=&\frac{2}{3}\sqrt{4\pi}\frac{1}{w^{(0)}_n}
\,\Big[q_n^{(0)}\langle\sin(q_n^{(0)}t)\Theta(t)\Theta(r_{\rm ex}-t)
\rangle-\langle\Delta^{(0)}(q_n^{(0)},r_{\rm ex},t)\rangle\nonumber \\
&-&\frac{1}{q_n^{(0)}}\lim_{\epsilon\rightarrow 0}
\langle\epsilon(t+\epsilon)^{-3}\sin(q_n^{(0)}t)\Theta(t)
\Theta(r_{\rm ex}-t)\rangle\Big], \\
\bar{c}^{(2\pm 2)}_{A_{xx}\,n}&=&-\sqrt{\frac{3}{2}}\bar{c}^{(20)}_{A_{xx}\,n}
=-\sqrt{\frac{2\pi}{15}}\frac{1}{w^{(2)}_n}
\,\Big[q_n^{(2)}\langle\sin(q_n^{(2)}t)\Theta(t)\Theta(r_{\rm ex}-t)\rangle 
+\langle\Delta^{(2)}(q_n^{(2)},r_{\rm ex},t)\rangle\Big];
\end{eqnarray}
for the BR geometric factor $\bar{A}_{xy}^{\rm(I,II)}$ by   
\begin{equation}
\bar{c}^{(2\pm 2)}_{A_{xy}\,n}
=\pm {\rm i}\sqrt{\frac{2\pi}{15}}\frac{1}{w^{(2)}_n}
\,\Big[q_n^{(2)}\langle\sin(q_n^{(2)}t)\Theta(t)\Theta(r_{\rm ex}-t)\rangle 
+\langle\Delta^{(2)}(q_n^{(2)},r_{\rm ex},t)\rangle\Big];
\end{equation}
and for the BR geometric factor $\bar{B}_{xy}^{\rm(I,II)}$ by   
\begin{equation}
\bar{c}^{(10)}_{B_{xy}\,n}
={\rm i}\sqrt{\frac{4\pi}{3}}\frac{1}{w^{(1)}_n}
\,\Big[q_n^{(1)}\langle\cos(q_n^{(1)}t)\Theta(t)\Theta(r_{\rm ex}-t)\rangle
+\langle\Delta^{(1)}(q_n^{(1)},r_{\rm ex},t)\rangle\Big].
\end{equation}
Here, the quantities $w_n^{(l)}$ are given by equation (16), and
the angular brackets denote the time averaging
\begin{equation}
\langle f(t)\rangle 
\equiv\frac{1}{\Delta t_1\Delta t_2}\int_0^{\Delta t_1}{\rm d}t_1
\int_T^{T+\Delta t_2}{\rm d}t_2\, f(t_2-t_1).
\end{equation}

Let us first evaluate the time averages 
$\langle\sin(qt)\Theta(t)\Theta(r_{\rm ex}-t)\rangle$ and 
$\langle\cos(qt)\Theta(t)\Theta(r_{\rm ex}-t)\rangle$.
A straightforward way of doing so is to integrate by parts 
in both $t_2$ and $t_1$, and to utilize the fact that 
$d\Theta(x)/dx=\delta(x)$:
\begin{eqnarray}
&&\int_0^{\Delta t_1}{\rm d}t_1\int_T^{T+\Delta t_2}{\rm d}t_2q
\sin[q(t_2-t_1)]
\Theta(t_2-t_1)\Theta(r_{\rm ex}-t_2+t_1) \nonumber \\ 
&&=\int_0^{\Delta t_1}\Big\{
-\cos[q(t_2-t_1)]\Theta(t_2-t_1)\Theta(r_{\rm ex}-t_2+t_1)
\Big|_{t_2=T}^{T+\Delta t_2}\nonumber \\
&&+\int_T^{T+\Delta t_2}\!\!\cos[q(t_2-t_1)]
[\delta (t_2-t_1)\Theta(r_{\rm ex}-t_2+t_1)-\Theta(t_2-t_1)\delta(r_{\rm ex}
-t_2+t_1)] {\rm d}t_2\Big\}{\rm d}t_1  \nonumber \\
&&=\frac{1}{q}\sin[q(T+\Delta t_2-t_1)]\Theta(T+\Delta t_2-t_1)
\Theta(r_{\rm ex}-T-\Delta t_2+t_1)\Big|_{t_1=0}^{\Delta t_1}\nonumber \\
&&-\frac{1}{q}\int_0^{\Delta t_1}\sin[q(T+\Delta t_2-t_1)]
\Theta(T+\Delta t_2-t_1)
\delta(r_{\rm ex}-T-\Delta t_2+t_1)\,{\rm d}t_1\nonumber \\
&&-\frac{1}{q}\sin[q(T-t_1)]\Theta(T-t_1)\Theta(r_{\rm ex}-T+t_1)
\Big|_{t_1=0}^{\Delta t_1}\nonumber \\
&&+\frac{1}{q}\int_0^{\Delta t_1}\sin[q(T-t_1)]
\Theta(T-t_1) \delta(r_{\rm ex}-T+t_1)\,{\rm d}t_1\nonumber \\
&&+\int_0^{\Delta t_1}\!\!\Big\{\Theta(t_1-T)\Theta(T+\Delta t_2-t_1)
-\cos(qr_{\rm ex})\Theta(r_{\rm ex}-T+t_1)\Theta(T+\Delta t_2-r_{\rm ex}
-t_1)\Big\} {\rm d}t_1.\nonumber\\ 
\end{eqnarray}
Here, terms with $\sin x\,\delta(x)$ were dropped immediately.
This gives for the time average 
$\langle\sin(qt)\Theta(t)\Theta(r_{\rm ex}{-}t)\rangle$: 
\begin{eqnarray}
&&\Delta t_1\Delta t_2\,q\,\langle\sin(qt)\Theta(t)\Theta(r_{\rm ex}-t)
\rangle \nonumber \\
&&=\frac{1}{q}\Big\{
\sin[q(T+\Delta t_2-\Delta t_1)]\Theta(T+\Delta t_2-\Delta t_1)
\Theta(r_{\rm ex}-T-\Delta t_2+\Delta t_1)\nonumber \\
&&-\sin[q(T+\Delta t_2)]\Theta(T+\Delta t_2) 
\Theta(r_{\rm ex}-T-\Delta t_2) \nonumber \\
&&-\sin[q(T-\Delta t_1)]\Theta(T-\Delta t_1)\Theta(r_{\rm ex}-T+\Delta t_1)
+\sin(qT)\Theta(T)\Theta(r_{\rm ex}-T)\nonumber \\
&&-\sin(qr_{\rm ex})\big
[\Theta(T+\Delta t_2-r_{\rm ex})\Theta(\Delta t_1-T-\Delta t_2 +r_{\rm ex})
-\Theta(T-r_{\rm ex})\Theta(\Delta t_1-T+r_{\rm ex})\big]\Big\}
\nonumber \\
&&-\cos(qr_{\rm ex})\Theta(\Delta t_1-T+r_{\rm ex})
\Theta(T+\Delta t_2-r_{\rm ex})
[{\rm min}(\Delta t_1, T+\Delta t_2-r_{\rm ex})-{\rm max}
(T-r_{\rm ex},0)]\nonumber \\
&&+\Theta(\Delta t_1-T)\Theta(T+\Delta t_2)
[{\rm min}(\Delta t_1, T+\Delta t_2) -{\rm max}(T,0)].
\end{eqnarray}

The time average $\langle\cos(qt)\Theta(t)\Theta(r_{\rm ex}{-}t)\rangle$,
calculated in a way similar to that for the time average (48), yields
\begin{eqnarray}
&&\Delta t_1\Delta t_2\,q\,\langle\cos(qt)\Theta(t)\Theta(r_{\rm ex}-t)
\rangle \nonumber \\
&&=\frac{1}{q}\Big\{
\cos[q(T+\Delta t_2-\Delta t_1)]\Theta(T+\Delta t_2-\Delta t_1)
\Theta(r_{\rm ex}-T-\Delta t_2+\Delta t_1)\nonumber \\
&&-\cos[q(T+\Delta t_2)]\Theta(T+\Delta t_2)\Theta(r_{\rm ex}-T-\Delta t_2) 
\nonumber \\
&&-\cos[q(T-\Delta t_1)]\Theta(T-\Delta t_1)\Theta(r_{\rm ex}-T+\Delta t_1)
+\cos(qT)\Theta(T)\Theta(r_{\rm ex}-T) \nonumber \\
&&-\cos(qr_{\rm ex})\big
[\Theta(T+\Delta t_2-r_{\rm ex})\Theta(\Delta t_1-T-\Delta t_2 +r_{\rm ex})
-\Theta(T-r_{\rm ex})\Theta(\Delta t_1-T+r_{\rm ex})\big]
\nonumber \\
&&+\Theta(T+\Delta t_2)\Theta(\Delta t_1-\Delta t_2-T)
-\Theta(T)\Theta(\Delta t_1-T)\Big\}+\sin(qr_{\rm ex})\nonumber \\      
&&\times\Theta(\Delta t_1-T+r_{\rm ex})
\Theta(T+\Delta t_2-r_{\rm ex})
[{\rm min}(\Delta t_1, T+\Delta t_2-r_{\rm ex})-{\rm max}(T-r_{\rm ex},0)].
\end{eqnarray}
The ambiguity that may arise in these expressions when the argument 
of the step function
vanishes is removed correctly by the definition $\Theta(0)=\case{1}{2}$. 
The correctness of the somewhat lengthy analytical expressions (48) and (49) 
was checked by performing the two-dimensional integration numerically
for cases with the time intervals $(0,\Delta t_1)$ and 
$(T,T+\Delta t_2)$ in all the possible logical relations of one to another. 

For the time averages $\langle\Delta^{(l)}(q,r_{\rm ex},t)\rangle$,
we need the time averages $\langle\delta(t{-}r_{\rm ex})\rangle$
and $\langle\delta'(t{-}r_{\rm ex})\rangle$, which are evaluated easily 
to give
\begin{eqnarray}
&&\Delta t_1\Delta t_2\,\langle\delta(t-r_{\rm ex})\rangle=
\Theta(\Delta t_1-T+ r_{\rm ex}) 
\Theta(T+\Delta t_2-r_{\rm ex})\nonumber \\
&&\times[{\rm min}(\Delta t_1,T+\Delta t_2-r_{\rm ex})  
-{\rm max}(T- r_{\rm ex},0)],\\
&&\Delta t_1\Delta t_2\,\langle\delta'(t-r_{\rm ex})\rangle=
\Theta(T+\Delta t_2-r_{\rm ex})\Theta(\Delta t_1-T-\Delta t_2+r_{\rm ex})  
\nonumber \\
&&-\Theta(T-r_{\rm ex})\Theta(\Delta t_1-T+r_{\rm ex}).  
\end{eqnarray}
These time averages vanish in the limit $r_{\rm ex}\rightarrow\infty$.
The $\epsilon\rightarrow 0$ term in equation (42) can be evaluated 
by taking the limit $\epsilon\rightarrow 0$ already after one
time integration, and the result is
\begin{eqnarray}
&&\frac{1}{q}\lim_{\epsilon\rightarrow 0}\langle\epsilon(t+\epsilon)^{-3}
\sin(qt)\Theta(t)\Theta(r_{\rm ex}-t)\rangle\nonumber\\
&&=\frac{\Theta(\Delta t_1-T)\Theta(T+\Delta t_2)}{2\Delta t_1\Delta t_2}
[{\rm min}(\Delta t_1,T+\Delta t_2)-{\rm max}(T,0)],
\end{eqnarray}
which equals one half of the $q$-independent term in (48), which, in turn,
equals the average $\langle\delta(t-r_{\rm ex})\rangle$ with $r_{\rm ex}=0$
of equation (50).

Using equations (48)--(52) 
and utilizing the fact that the quantities $q^{(l)}_n r_{\rm ex}$ are the 
roots of the 
spherical Bessel functions $j_l(x)$ and thus $j_l(q^{(l)}_n r_{\rm ex})=0$,
the time averaged Fourier--Bessel coefficients (42)--(45)
are evaluated finally as follows: 
\begin{eqnarray}
\bar{c}^{(00)}_{A_{xx}\,n}&=&\frac{2}{3}\frac{\sqrt{4\pi}}
{\Delta t_1\Delta t_2w^{(0)}_nq_n^{(0)}}\nonumber \\
&&\times\Big\{
\sin[q_n^{(0)}(T+\Delta t_2-\Delta t_1)]\Theta(T+\Delta t_2-\Delta t_1)
\Theta(r_{\rm ex}-T-\Delta t_2+\Delta t_1)\nonumber \\
&&-\sin[q_n^{(0)}(T+\Delta t_2)]\Theta(T+\Delta t_2) 
\Theta(r_{\rm ex}-T-\Delta t_2) \nonumber \\
&&-\sin[q_n^{(0)}(T-\Delta t_1)]\Theta(T-\Delta t_1)\Theta(r_{\rm ex}
-T+\Delta t_1)
+\sin(q_n^{(0)}T)\Theta(T)\Theta(r_{\rm ex}-T)\nonumber \\
&&-\case{1}{2}q_n^{(0)}\Theta(\Delta t_1-T)\Theta(T+\Delta t_2)
[{\rm min}(\Delta t_1, T+\Delta t_2) -{\rm max}(T,0)]\Big\},
\end{eqnarray}
\begin{eqnarray}
\bar{c}^{(2\pm 2)}_{A_{xx}\,n}&=&-\sqrt{\frac{3}{2}}\bar{c}^{(20)}_{A_{xx}\,n}
=\pm{\rm i}\bar{c}^{(2\pm 2)}_{A_{xy}\,n}  
=-\sqrt{\frac{2\pi}{15}}\frac{1}{\Delta t_1\Delta t_2w^{(2)}_nq_n^{(2)}}
\nonumber \\
&&\times\Big\{
\sin[q_n^{(2)}(T+\Delta t_2-\Delta t_1)]\Theta(T+\Delta t_2-\Delta t_1)
\Theta(r_{\rm ex}-T-\Delta t_2+\Delta t_1)\nonumber \\
&&-\sin[q_n^{(2)}(T+\Delta t_2)]\Theta(T+\Delta t_2) 
\Theta(r_{\rm ex}-T-\Delta t_2) \nonumber \\
&&-\sin[q_n^{(2)}(T-\Delta t_1)]\Theta(T-\Delta t_1)\Theta(r_{\rm ex}
-T+\Delta t_1)
+\sin(q_n^{(2)}T)\Theta(T)\Theta(r_{\rm ex}-T)\nonumber \\
&&-\sin(q_n^{(2)}r_{\rm ex})\{
\Theta(T+\Delta t_2-r_{\rm ex})\Theta(\Delta t_1-T-\Delta t_2 +r_{\rm ex})
\nonumber\\
&&-\Theta(T-r_{\rm ex})\Theta(\Delta t_1-T+r_{\rm ex})
+\Theta(\Delta t_1-T+ r_{\rm ex})\Theta(T+\Delta t_2-r_{\rm ex})\nonumber\\
&&\times r^{-1}_{\rm ex}[{\rm min}(\Delta t_1,T+\Delta t_2-r_{\rm ex})
-{\rm max}(T- r_{\rm ex},0)]\}\nonumber \\
&&+q_n^{(2)}\Theta(\Delta t_1-T)\Theta(T+\Delta t_2)
[{\rm min}(\Delta t_1, T+\Delta t_2) -{\rm max}(T,0)]\Big\},
\end{eqnarray}
\begin{eqnarray}
\bar{c}^{(10)}_{B_{xy}\,n}
&=&{\rm i}\sqrt{\frac{4\pi}{3}}\frac{1}{\Delta t_1\Delta t_2w^{(1)}_n
q_n^{(1)}}\nonumber \\
&&\times \Big\{
\cos[q_n^{(1)}(T+\Delta t_2-\Delta t_1)]\Theta(T+\Delta t_2-\Delta t_1)
\Theta(r_{\rm ex}-T-\Delta t_2+\Delta t_1)\nonumber \\
&&-\cos[q_n^{(1)}(T+\Delta t_2)]\Theta(T+\Delta t_2)
\Theta(r_{\rm ex}-T-\Delta t_2) 
\nonumber \\
&&-\cos[q_n^{(1)}(T-\Delta t_1)]\Theta(T-\Delta t_1)
\Theta(r_{\rm ex}-T+\Delta t_1)
+\cos(q_n^{(1)}T)\Theta(T)\Theta(r_{\rm ex}-T) \nonumber \\
&&-\cos(q_n^{(1)}r_{\rm ex})
[\Theta(T+\Delta t_2-r_{\rm ex})\Theta(\Delta t_1-T-\Delta t_2 +r_{\rm ex})
\nonumber \\
&&-\Theta(T-r_{\rm ex})\Theta(\Delta t_1-T+r_{\rm ex})]
+\Theta(T+\Delta t_2)\Theta(\Delta t_1-\Delta t_2-T)\nonumber \\
&&-\Theta(T)\Theta(\Delta t_1-T)\Big\}.      
\end{eqnarray}

Equations (25) and (53)--(55) furnish a general solution to the
problem of finding Fourier--Bessel expansions of the representative 
BR geometric factors (6)--(8) with space regions specified by the multipoles
(24) of their Fourier transforms. Obviously, the formalism developed can 
easily be used to give Fourier--Bessel expansions of all possible
BR geometric factors, and not only the representatives (6)--(8), as long
as their space regions have well-behaved multipole expansions.

\begin{flushleft} 
\bf 3. BR geometric factors with spherical space regions
\end{flushleft}
\begin{flushleft} 
\it 3.1 Fourier--Bessel expansions
\end{flushleft}
Formula (25) gives a Fourier--Bessel expansion of the BR geometric 
factor $\bar{C}^{(\rm I,II)}_U$ for the general case when the uniform
densities 
$\rho_k(\bold{r}_k)$ as well as the function $U(t,\bold{r})$ are not 
spherically symmetric functions of space coordinates. 
Let us assume now
that the densities are spherically symmetric with radii $R_k$, 
$\rho_k(\bold{r}_k)=\rho_k(r_k)=(3/4\pi R_k^3)\Theta(R_k-r_k)$.
Such an assumption 
should not entail any serious loss in generality as space regions of 
practical relevance can be approximated by regions of spherical shape,
but the main reason for considering spherical space regions is that
it simplifies considerably the formulation of the problem and the
actual calculations.
With spherical space regions,  
only the $l_1=l_2=0$ terms contribute in the general formula (25), 
Substituting further 
$\tilde{\rho}^{(k)}_{00}(q)=(4\pi)^{1/2}3j_1(qR_k)/qR_k$
for the Fourier transforms of the spherical uniform densities $\rho_k(r_k)$, 
equation (25)
simplifies to
\begin{equation}
\bar{C}^{\rm(I,II)}_U=\frac{9}{R_1R_2}\sum_{lm}\sum_{n=1}^{\infty}
\frac{\bar{c}^{(lm)}_{U\,n}}
{(q_n^{(l)})^2}j_1\big(q_n^{(l)}R_1\big)j_1\big(q_n^{(l)}R_2\big)
j_l\big(q_n^{(l)}R\big){\rm i}^lY_{lm}\big(\hat{\bold{R}}\big).
\end{equation}
But more importantly, this assumption 
allows an alternative and simpler formulation based on 
a Fourier--Bessel expansion
of one of the spherically symmetric densities, say $\rho_1(r_1)$, instead
of the one based on the Fourier--Bessel expansion of the function 
$U(t,\bold{r})$. 

Let us then expand the uniform density $\rho_1(r_1)$ as 
a Fourier--Bessel series in the 
spherical Bessel functions $j_0(q_nr_1)$ in a range $0\le r_1<r_{\rm ex}$: 
\begin{equation}
\rho_1(r_1)=\sum_{n=1}^{\infty}c_nj_0(q_nr_1).
\end{equation}
Here, $q_n=n\pi/r_{\rm ex}$, and the coefficients $c_n$ are given by
\begin{equation}
c_n=\frac{2}{r_{\rm ex}}\left(\frac{n\pi}{r_{\rm ex}}\right)^2
\int_0^{r_{\rm ex}}\rho_1(r_1)j_0\left(\frac{n\pi}{r_{\rm ex}}r_1\right)
r_1^2\,{\rm d}r_1 
=\frac{3n}{2r_{\rm ex}^2{R_1}'}\,j_1\left(\frac{n\pi}{r_{\rm ex}}{R_1}'
\right), 
\end{equation}
where ${R_1}'={\rm min}(R_1,r_{\rm ex})$, with $R_1$ the density's radius.
Since the value of the density $\rho_1(r_1)$ is needed in the multiple
integral (9) that defines the BR geometric factor $\bar{C}^{(\rm I,II)}_U$
only when
\begin{equation}
r_1=|\bold{R}+\bold{r}_2-\bold{r}|\le r_{1\,{\rm max}}
= R+R_2+ {\rm max}(T+\Delta t_2,0),
\end{equation}
where $R$ is the separation of the centres of the two densities, and $R_2$ 
and ${\rm max}(T+\Delta t_2,0)$ are respectively the radii beyond which the 
density 
$\rho_2(r_2)$ and the function $U(t,\bold{r})$ vanish, 
the expansion radius $r_{\rm  ex}$ should satisfy
the relation $r_{\rm  ex}\ge r_{1\,\rm max}$.
Substituting then the expansion (57) with an expansion radius $r_{\rm ex}
\ge r_{1\,\rm max}$ in equation (9), and utilizing the identity (18) 
for $j_0(q_nr_1)$,
\begin{equation}
j_0(q_nr_1)=\frac{1}{4\pi}\int\exp({\rm i}\bold{q}_n{\bf\cdot}\bold{r}_1)\,
{\rm d}\hat{\bold{q}}_n
\end{equation} 
with $\bold{r}_1 =\bold{R}+\bold{r}_2-\bold{r}$,
one obtains for the BR geometric factor $\bar{C}^{(\rm I,II)}_U$ 
a Fourier--Bessel expansion: 
\begin{equation}
\bar{C}^{(\rm I,II)}_U =\frac{1}{4\pi}\sum_{n=1}^{\infty}c_n
\tilde{\rho}_2(q_n)\int\bar{\tilde{U}}(-\bold{q}_n) 
\exp({\rm i}\bold{q}_n{\bf\cdot}\bold{R})\,
{\rm d}\hat{\bold{q}}_n,
\end{equation}
Here, 
\begin{equation}
\tilde{\rho}_2(q_n)=4\pi\int_0^{\infty}\rho_2(r_2)j_0(q_nr_2)r_2^2{\rm d}r_2
=3\frac{j_1(q_nR_2)}{q_nR_2}
\end{equation}
is the Fourier transform of the uniform density $\rho_2(r_2)$ with a radius
$R_2$, and $\bar{\tilde{U}}(\bold{q})$ is the time averaged Fourier transform 
of the function $U(t,\bold{r})$: 
\begin{equation}
\bar{\tilde{U}}(\bold{q})=
\frac{1}{\Delta t_1 \Delta t_2}
\int_0^{\Delta t_1}{\rm d}t_1\int_{T}^{T+\Delta t_2}{\rm d}t_2\,  
\int {\rm d}\bold{r}\, U(t,\bold{r})\,{\rm exp}({\rm i}
\bold{q}{\bf\cdot}\bold{r}).
\end{equation}
With the multipole expansion (13) of $U(t,\bold{r})$ and a further use of 
the identity (18), 
the time averaged Fourier transform (64) can be written also as a multipole 
sum
\begin{equation}
\bar{\tilde{U}}(\bold{q})=\sum_{lm}\bar{\tilde{U}}_{lm}(q)
{\rm i}^l Y_{lm}(\hat{\bold{q}}),
\end{equation}
where
\begin{equation}
\bar{\tilde{U}}_{lm}(q)=\frac{4\pi{\rm i}^l}{\Delta t_1 \Delta t_2}
\int_0^{\Delta t_1}{\rm d}t_1\int_{T}^{T+\Delta t_2}{\rm d}t_2\, 
\int_0^{\infty}r^2 {\rm d}r\, U_{lm}(t,r)j_l(qr).
\end{equation}
Using the multipole expansion (64) and again the identity (18), the 
Fourier--Bessel expansion (61) of $\bar{C}^{(\rm I,II)}_U$                
now takes the form of a multipole sum:
\begin{equation}
\bar{C}^{(\rm I,II)}_U =\sum_{lm}\sum_{n=1}^{\infty}c_n
\tilde{\rho}_2(q_n) \bar{\tilde{U}}_{lm}(q_n)j_l(q_nR)Y_{lm}(\hat{\bold{R}}). 
\end{equation}
Substituting here for $c_n$ and $\tilde{\rho}_2(q_n)$ from equations (58)
and (62), respectively, and writing $q_n$ explicitly as $n\pi/r_{\rm ex}$, 
one obtains finally
\begin{mathletters}
\begin{equation}
\bar{C}^{(\rm I,II)}_U =\frac{9}{2\pi r_{\rm ex}{R_1}'R_2}
\sum_{lm}\sum_{n=1}^{\infty}
j_1\left(\frac{n\pi}{r_{\rm ex}}{R_1}'\right)
j_1\left(\frac{n\pi}{r_{\rm ex}}R_2\right)
\bar{\tilde{U}}_{lm}\left(\frac{n\pi}{r_{\rm ex}}\right)
j_l\left(\frac{n\pi}{r_{\rm ex}}R\right)Y_{lm}(\hat{\bold{R}}), 
\label{a}
\end{equation}
\begin{equation}
{R_1}'={\rm min}(R_1, r_{\rm ex}),\;\;\;\;\;r_{\rm ex}\ge R+R_2
+{\rm max}(T+\Delta t_2, 0).
\label{b} 
\end{equation}
\end{mathletters}
\indent Note that while different spherical Bessel function roots 
$q_n^{(l)}r_{\rm ex}$ and associated weights $w_n^{(l)}$ are  
required with the different multipolarities $l$ of the function 
$U(t,\bold{r})$ 
in the formula (56) based on the Fourier--Bessel expansion of this function, 
here only the simple quantities
$q^{(0)}_n=n\pi/r_{\rm ex}$ and $w^{(0)}_n=r_{\rm ex}^3/2(n\pi)^2$  are 
used for all these multipolarities. For this practical reason, 
when one of the densities is a spherically symmetric function,
the formulation based on the Fourier--Bessel expansion (57)  
is preferable to an application of the general formula (25) to such a case. 
Moreover, because the infinite-radius Fourier transform of the 
function $U(t,\bold{r})$ is now used, rather than the finite-radius 
transforms (33)--(35), the expressions for the time-averaged multipoles (65) 
of the Fourier transform will be simpler than those 
required in (42)--(45) for the coefficients 
$\bar{c}^{(lm)}_{U\,n}$  needed in the general formulation. 
It follows from equations (15), (21), (42)--(45), (50)--(52) and (65) that
the quantities $\bar{\tilde{U}}_{lm}(n\pi/r_{\rm ex})$ needed in (67) 
for the BR geometric factor $\bar{A}_{xx}^{\rm(I,II)}$ are given by
\begin{eqnarray}
{\bar{\tilde U}^{(A_{xx})}_{00}}(q)&=&
(4\pi)^{3/2}\frac{2}{3}\Big[q\,\langle\sin(qt)\Theta(t)\rangle
-\case{3}{2}\langle\delta(t)\rangle\Big], \\
\bar{\tilde U}^{(A_{xx})}_{2\pm 2}(q)
&=&-\sqrt{\frac{3}{2}}\,
\bar{\tilde U}^{(A_{xx})}_{20}(q) 
=4\pi\sqrt{\frac{2\pi}{15}}\,q
\,\langle\sin(qt)\Theta(t)\rangle; 
\end{eqnarray}
for the BR geometric factor $\bar{A}_{xy}^{\rm(I,II)}$ by   
\begin{equation}
\bar{\tilde U}^{(A_{xy})}_{2\pm 2}(q)
=\mp{\rm i}4\pi \sqrt{\frac{2\pi}{15}}q
\,\langle\sin(qt)\Theta(t)\rangle; 
\end{equation}
and for the BR geometric factor $\bar{B}_{xy}^{\rm(I,II)}$ by   
\begin{equation}
\bar{\tilde U}^{(B_{xy})}_{10} (q)
=-4\pi\sqrt{\frac{4\pi}{3}}q\,
\langle\cos(qt)\Theta(t)\rangle.
\end{equation}
Here, the time averages $\langle\sin(qt)\Theta(t)\rangle$ 
and $\langle\cos(qt)\Theta(t)\rangle$ are the limits 
$r_{\rm ex}\rightarrow \infty$
of the finite-radius time averages (48) and (49):
\begin{eqnarray}
&&\Delta t_1\Delta t_2\,q\,\langle\sin(qt)\Theta(t)\rangle \nonumber \\
&&=\frac{1}{q}\Big\{
\sin[q(T+\Delta t_2-\Delta t_1)]\Theta(T+\Delta t_2-\Delta t_1)
-\sin[q(T+\Delta t_2)]\Theta(T+\Delta t_2) \nonumber \\
&&-\sin[q(T-\Delta t_1)]\Theta(T-\Delta t_1)+\sin(qT)\Theta(T)\Big\}
\nonumber \\
&&+\Theta(\Delta t_1-T)\Theta(T+\Delta t_2)
[{\rm min}(\Delta t_1, T+\Delta t_2) -{\rm max}(T,0)], \\
&&\Delta t_1\Delta t_2\,q\,\langle\cos(qt)\Theta(t)\rangle \nonumber \\
&&=\frac{1}{q}\Big\{
\cos[q(T+\Delta t_2-\Delta t_1)]\Theta(T+\Delta t_2-\Delta t_1)
-\cos[q(T+\Delta t_2)]\Theta(T+\Delta t_2) \nonumber \\
&&-\cos[q(T-\Delta t_1)]\Theta(T-\Delta t_1)+\cos(qT)\Theta(T)
\nonumber \\
&&+\Theta(T+\Delta t_2)\Theta(\Delta t_1-\Delta t_2-T)
  -\Theta(T)\Theta(\Delta t_1-T)\Big\},
\end{eqnarray}
and the quantity $\langle\delta(t)\rangle$ in (68) is the time average (50) 
with $r_{\rm ex}=0$,
which also happens to equal the $q$-independent term in (72) and twice the 
$\epsilon\rightarrow 0$ term of equation (52). 

The greatly simplified (when compared with those of the general formulation) 
equations (67)--(73) give the Fourier--Bessel expansions of the 
representative BR geometric factors (6)--(8) with spherical space regions; 
it is gratifying to note that it was possible to express the terms of these
expansions in terms of elementary functions.
In fact, it turns out that the BR geometric factors with
spherical space regions can be evaluated in closed form, which will 
provide a means of testing the accuracy of the Fourier--Bessel results, 
but this is most 
easily accomplished starting with the standard Fourier-integral 
treatment of the multiple folding integrals involved.

\begin{flushleft} 
\it 3.2 Closed-form evaluation
\end{flushleft}
Using the standard Fourier-integral method of calculating
multiple folding integrals, a BR geometric factor $\bar{C}^{\rm(I,II)}_U$ 
with spherical space regions $\rho_k(r_k)$,
\begin{equation}
\bar{C}^{(\rm I,II)}_U =\frac{1}{\Delta t_1 \Delta t_2}
\int_0^{\Delta t_1}{\rm d}t_1\int_{T}^{T+\Delta t_2}{\rm d}t_2\, 
\int\rho_1(r_1){\rm d}\bold{r}_1\int\rho_2(r_2){\rm d}\bold{r}_2\,
U(t,\bold{r}), 
\end{equation}
can be written as a multipole expansion, the multipoles of which
are Fourier integrals:
\begin{equation}
\bar{C}^{(\rm I,II)}_U=\frac{4\pi}{(2\pi)^3}\sum_{lm}\int_0^{\infty}
\tilde{\rho}_1(q)\tilde{\rho}_2(q)\bar{\tilde{U}}_{lm}(q)j_l(qR)q^2{\rm d}q
Y_{lm}(\hat{\bold{R}}). 
\end{equation}
Here, $\bar{\tilde{U}}_{lm}(q)$ are the multipoles (65) of the 
time-averaged Fourier transform of the function $U(t,\bold{r})$, 
given by equations (68)--(73), and
\begin{equation}
\tilde{\rho}_k(q)=4\pi\int_0^{\infty}\rho_k(r_k)j_0(qr_k)r_k^2{\rm d}r_k
=3\frac{j_1(qR_k)}{qR_k}
\end{equation}
are the Fourier transforms of the spherically symmetric uniform densities
$\rho_k(r_k)$, which have radii $R_k$ and are normalized to unit volume.
This result follows from the convolution theorem 
(see, for example, \cite{Arf}) according to which the 
Fourier transform of a folding integral, like the one in equation (74), 
is the
product of the Fourier transforms of the functions appearing in the folding
integral; equation (75) is simply the time average  of the Fourier 
transformation
of the folding integral from the momentum space back to the configuration
space. In general, Fourier integrals of the type of those in equation (75) 
have to be 
evaluated numerically, and the great advantage of the Fourier--Bessel
formulation is that it replaces such numerical integration by the
analytical method of a series expansion. However, in the case
of spherical space regions with the Fourier transforms (76), and with the
time-averaged Fourier transforms (68)--(73) of the functions $U(t,\bold{r})$,
the Fourier integrals in (75) can in fact be evaluated in closed 
form, and we present such evaluation in this section. 
 
The quantities $\bar{\tilde{U}}_{lm}(q)$ in (75), as given by 
equations (68)--(73),  are linear combinations of terms of the form 
$\tau j_0(\tau q)$ and a $q$-independent term for the multipolarities 
$l=0,2$, and of terms of the form
$\tau j_{-1}(\tau q)$ and a $1/q$ term for the multipolarity $l=1$, while
the Fourier transforms (76) of the densities are of the form $j_1(aq)/aq$.
We define therefore the integrals
\begin{equation}
{\rm ji}_4(n;l_1,l_2,l_3,l_4;\alpha,\beta,\gamma,\delta)
=\int_0^{\infty}j_{l_1}(\alpha x)j_{l_2}(\beta x)
j_{l_3}(\gamma x)j_{l_4}(\delta x)\,\frac{{\rm d}x}{x^n}, 
\end{equation}
the evaluation of which
is required in (75) for the parameter values (i)
$n=0$, $l_1=l_2=1$, $l_3=0$, $l_4=0,2$; (ii) $n=0$, $l_1=l_2=1$, $l_3=-1$, 
$l_4=1$; and, on account of the $1/q$ term in (73), (iii) $n=1$,
$l_1=l_2=1$, $l_3=0$ (with $\gamma=0$), $l_4=1$.

In principle, it should be possible to evaluate the integral 
${\rm ji}_4(n;l_1,l_2,l_3,l_4;\alpha,\beta,\gamma,\delta)$
in closed form for all the integer values $n$ and $l_1,l_2,l_3,l_4$
for which the integral exists, as the integrand can be written as 
a sum of a finite number of terms, with each term having the form 
$a\sin(bx)/x^k$ or $a\cos(bx)/x^k$, where $k$ is an integer, and the 
indefinite integrals of such terms
can be done in terms of the sine or cosine integrals. In practice, however,
such a procedure is prohibitively lengthy for all but
very small values of the parameters $n,l_1,l_2,l_3,l_4$: for example, 
the integrand of
the integral ${\rm ji}_4(0;1,1,0,2;\alpha,\beta,\gamma,\delta)$ has
already 96 terms of the above mentioned form with $k$ ranging from 4 to 8,
and the indefinite integral of each of these terms generates in turn 
$k$ new terms, giving in total 572 terms. Furthermore, complications 
arise when the limit 
$x\rightarrow 0$ (or $x\rightarrow\infty$, depending on the definition of 
the sine integral) is taken in the sine integrals, which have arguments of 
the form $ax$, as the signs of the parameters $a$, which  are linear 
combinations of the parameters $\alpha$, $\beta$, $\gamma$ and $\delta$, 
must be determined suitably. 
Remarkably, the computing system {\it Mathematica}\cite{Wolfram} 
is able to perform the definite integrations in the integrals 
${\rm ji}_4(n;l_1,l_2,l_3,l_4;\alpha,\beta,\gamma,\delta)$ required directly. 
After considerable simplifications, and using an economic  
way of writing linear combinations that contain terms with permuting signs 
by means of the definitions 
\begin{equation}
\alpha_n=\alpha,\;\;\;\;\beta_n=(-1)^n\beta,\;\;\;\;
\gamma_n=(-1)^{[n/2]}\gamma,\;\;\;\;\delta_n=(-1)^{[n/4]}\delta,
\;\;\;\;\delta_n'=(-1)^{[n/2]}\delta,
\end{equation}
where $[x]$ is the integer part of $x$, the results are as follows: 
\begin{eqnarray}
&&{\rm ji}_4(0;1,1,0,0;\alpha,\beta,\gamma,\delta)  
=\frac{\pi}{1920\alpha\beta}\sum_{n=0}^7\frac{|\alpha_n+\beta_n+\gamma_n
+\delta_n|^3}{\alpha_n\beta_n\gamma_n\delta_n}
\nonumber \\
&&\times [4\alpha_n^2+(4\beta_n-\gamma_n-\delta_n)(-3\alpha_n+\beta_n
+\gamma_n+\delta_n)];
\end{eqnarray}
\begin{eqnarray}
&&{\rm ji}_4(0;1,1,0,2;\alpha,\beta,\gamma,\delta) 
=-\frac{\pi}{26\,880\alpha\beta\delta^2}\sum_{n=0}^7\frac{|\alpha_n+\beta_n
+\gamma_n+\delta_n|^3}{\alpha_n\beta_n\gamma_n\delta_n}
\nonumber \\
&&\times\{(\alpha_n+\beta_n+\gamma_n)(\alpha_n+\beta_n+\gamma_n-3\delta_n)
[6\alpha_n^2+(6\beta_n-\gamma_n)(-5\alpha_n+\beta_n+\gamma_n)]\nonumber \\
&&+[8(\alpha_n^2-12\alpha_n\beta_n+\beta_n^2)+9(\alpha_n+\beta_n)\gamma_n
+\gamma_n^2]\delta_n^2
+[24(\alpha_n+\beta_n)-11\gamma_n]\delta_n^3-8\delta_n^4\};
\end{eqnarray}
\begin{eqnarray}
&&{\rm ji}_4(0;1,1,-1,1;\alpha,\beta,\gamma,\delta) 
=-\frac{\pi}{11\,520\alpha\beta\gamma\delta}\sum_{n=0}^7\frac{|\alpha_n
+\beta_n+\gamma_n+\delta_n|^3}{\alpha_n\beta_n\delta_n}
\nonumber \\
&&\times\{5\alpha_n^3-3\alpha_n^2(5\beta_n-3\gamma_n+5\delta_n) 
+(-3\alpha_n+\beta_n+\gamma_n+\delta_n)\nonumber \\
&&\times[5\beta_n^2+(\gamma_n-5\delta_n)(4\beta_n-\gamma_n-\delta_n)]\}.
\end{eqnarray}
Here, it is assumed that the parameters $\alpha$, $\beta$, $\gamma$ and
$\delta$ are all nonzero;   
the required integrals that have some of these parameters equal to zero
were evaluated separately:
\begin{equation}
{\rm ji}_4(0;1,1,0,0;\alpha,\beta,0,0)=\frac{\pi}{12\alpha\beta}
\sum_{n=0}^1\frac{|\alpha_n+\beta_n|}{\alpha_n\beta_n}(\alpha^2_n-\alpha_n
\beta_n+\beta_n^2);
\end{equation}
\begin{eqnarray}
&&{\rm ji}_4(0;1,1,0,0;\alpha,\beta,\gamma,0)  
=\frac{\pi}{192\alpha\beta}\sum_{n=0}^3\frac{(\alpha_n+\beta_n+\gamma_n)
|\alpha_n+\beta_n+\gamma_n|}{\alpha_n\beta_n\gamma_n}
\nonumber \\
&&\times [3(\alpha_n-\beta_n)^2+2(\alpha_n+\beta_n)\gamma_n-\gamma_n^2];
\end{eqnarray}
\begin{eqnarray}
&&{\rm ji}_4(0;1,1,0,2;\alpha,\beta,0,\delta)
=-\frac{\pi}{384\alpha\beta\delta^2}\sum_{n=0}^3\frac{(\alpha_n+\beta_n
+\delta_n')|\alpha_n+\beta_n+\delta_n'|}{\alpha_n\beta_n\delta_n'}
\nonumber \\
&&\times(\alpha_n+\beta_n-\delta_n')^2(\alpha_n^2-4\alpha_n\beta_n
+\beta_n^2-{\delta'_n}^2);
\end{eqnarray}
\begin{eqnarray}
&&{\rm ji}_4(1;1,1,0,1;\alpha,\beta,0,\delta)
=-\frac{\pi}{1152\alpha\beta\delta}\sum_{n=0}^3\frac{|\alpha_n+\beta_n
+\delta_n'|^3}{\alpha_n\beta_n\delta_n'}
\nonumber \\
&&\times[\alpha_n^3-3\alpha_n(\beta_n^2-4\beta_n\delta_n'+{\delta'_n}^2)
+(\beta_n+\delta_n')
(-3\alpha_n^2+\beta_n^2-4\beta_n\delta_n'+{\delta'_n}^2)].
\end{eqnarray}
Here, the definition $j_0(0)=1$ is assumed in the integrals (77);
as $\lim_{x\rightarrow 0}j_l(x)=0$ when $l>0$, the integrals 
${\rm ji}_4(n;1,1,l_3,l_4;\alpha,\beta,\gamma,0)$ with $l_4>0$ vanish.
The integral ${\rm ji}_4(0;1,1,0,0;\alpha,\beta,0,\delta)$ that is required
also is given already by the integral (83):
\begin{equation}
{\rm ji}_4(0;1,1,0,0;\alpha,\beta,0,\delta)
={\rm ji}_4(0;1,1,0,0;\alpha,\beta,\delta,0).
\end{equation}

Using equations (68)--(73), (75) and (76), the integrals (79)--(86), 
and the definitions
\begin{eqnarray}
&&a_{00}^{(A_{xx})}=(4\pi)^{3/2}\,\frac{2}{3},\;\;\;\; 
a_{2\pm 1}^{(A_{xx})}=0,\;\;\;\; 
a_{2\pm2}^{(A_{xx})}=
-\sqrt{\frac{3}{2}}\,a_{20}^{(A_{xx})}=4\pi\sqrt{\frac{2\pi}{15}},\\
&&a_{20}^{(A_{xy})}=a_{2\pm 1}^{(A_{xy})}=0,\;\;\;\; 
a_{2\pm 2}^{(A_{xy})}=\mp{\rm i}4\pi\sqrt{\frac{2\pi}{15}},\;\;\;\;
b_{10}^{(B_{xy})}=-4\pi\sqrt{\frac{4\pi}{3}};
\end{eqnarray}
\begin{equation}
\tau_0=0,\;\;\;\;\tau_1=T+\Delta t_2-\Delta t_1,\;\;\;\;\tau_2=
T+\Delta t_2,\;\;\;\;\tau_3=T,\;\;\;\;\tau_4=T-\Delta t_1;
\end{equation}
\begin{eqnarray}
&&g_0^{(0)}
=-\frac{\Theta(-\tau_4)\Theta(\tau_2)}{2\Delta t_1\Delta t_2}
[{\rm min}(\Delta t_1, \tau_2) -{\rm max}(\tau_3,0)], \\
&&g_0^{(1)}=\frac{1}{\Delta t_1\Delta t_2}
[\Theta(\tau_2)\Theta(-\tau_1)-\Theta(\tau_3)\Theta(-\tau_4)],\\
&&g_0^{(2)}=
\frac{\Theta(-\tau_4)\Theta(\tau_2)} {\Delta t_1\Delta t_2}
[{\rm min}(\Delta t_1, \tau_2) -{\rm max}(\tau_3,0)], \\
&&g_i^{(l)}=(-1)^{i+1}\frac{\Theta(\tau_i)}{\Delta t_1\Delta t_2}
[\tau_i+{\rm zr}(\tau_i)\delta_{1l}],\;\;l=0,1,2,\;\;i=1,2,3,4,
\end{eqnarray}
where 
\begin{equation}
{\rm zr}(x)=\left\{\begin{array}{ccc}1 & {\rm for}&x =  0\\
                                     0 & {\rm for}&x\ne 0
                   \end{array}\right., 
\end{equation}                   
and the definition $\Theta(0)=\case{1}{2}$ is used again, 
we obtain finally the following closed-form expressions for the 
representative BR geometric factors (6)--(8) with spherical space regions:
\begin{eqnarray}
&&\bar{A}^{\rm(I,II)}_{xx}
=\frac{4\pi}{(2\pi)^3}\frac{9}{R_1R_2}
\sum_{\scriptstyle l=0,2\atop \scriptstyle m}a_{lm}^{(A_{xx})}
\sum_{i=0}^4g_i^{(l)}\,
{\rm ji}_4(0;1,1,0,l;R_1,R_2,\tau_i,R) Y_{lm}(\hat{\bold{R}}),\\
&&\bar{A}^{\rm(I,II)}_{xy}
=\frac{4\pi}{(2\pi)^3}\frac{9}{R_1R_2}
\sum_{m}a_{2m}^{(A_{xy})}\sum_{i=0}^4g_i^{(2)}\,
{\rm ji}_4(0;1,1,0,2;R_1,R_2,\tau_i,R) Y_{2m}(\hat{\bold{R}}),\\
&&\bar{B}^{\rm(I,II)}_{xy}
=\frac{4\pi}{(2\pi)^3}\frac{9}{R_1R_2}
b_{10}^{(B_{xy})}\sum_{i=0}^4g_i^{(1)}\,
{\rm ji}_4[{\rm zr}(\tau_i);1,1,{\rm zr}(\tau_i){-}1,1;R_1,R_2,\tau_i,R] 
Y_{10}(\hat{\bold{R}}).
\end{eqnarray}

\begin{flushleft} 
\bf 3. Numerical results and concluding remarks
\end{flushleft}
The convergence properties of the Fourier--Bessel expansions  
were examined in numerical calculations of some representative examples
of BR geometric coefficients with spherical space regions, using the 
formula (67) with the Fourier--Bessel coefficients given by equations 
(68)--(73)
and an expansion radius $r_{\rm ex}=R+R_2+{\rm max}(T+\Delta t_2,0)$.

For the BR geometric factors to have an appreciable magnitude, it is obvious 
that, in a system of units where the speed of light $c=1$,
the separation in space and time of the space-time regions must not 
be much greater than the dimensions of the space-time regions themselves. 
The calculations were done using an unspecified unit of length;
choosing the millimetre as the unit of length, for example, the unit of time 
equals approximately  $3.33 \times 10^{-12}$ s in  
a system of units with $c=1$. This illustrates the fact that, on a realistic 
laboratory scale, the time intervals corresponding to even relatively large 
distances are very short, but we leave aside the question of how 
field measurements occupying and/or separated by such short time intervals
can be realized.

The numerical values of the representative BR geometric factors 
(6)--(8) for various spherical space-time regions with dimensions of the 
order of unity
and similar or smaller space-time separations, in a $c=1$ system of units, 
are collected in table 1. 
As the quantities that are required in the field commutation relations
are the values of the differences 
$\bar{C}^{\rm(I,II)}_U-\bar{C}^{\rm(II,I)}_U$,
while some BR spring constants require the value of the sum
$\bar{C}^{\rm(I,II)}_U+\bar{C}^{\rm(II,I)}_U$,           
the ``reverse" geometric factor $\bar{C}^{\rm(II,I)}_U$ was calculated 
together with a given geometric factor $\bar{C}^{\rm(I,II)}_U$. This was 
done by interchanging the radii $R_1$ and $R_2$ of the 
spherical space regions and changing their relative displacement $\bold{R}$ 
to
$-\bold{R}$ (i.e., changing the polar angles $\theta$ and $\phi$ of $\bold{R}$
to $\pi-\theta$ and $\phi+\pi$), together with interchanging the time 
intervals $\Delta t_1$ and $\Delta t_2$ and changing their separation $T$ 
to $-T$. 
BR geometric factors $\bar{A}^{\rm(I,I)}_{xx}$ with fully coinciding 
space-time regions were calculated also as they are required for
some of the BR spring constants.
When the sphere separation $R=0$, only the monopole ($l=0$) term 
contributes in the expansion (67) because $j_l(q_nR)\rightarrow 0$ for $l>0$ 
as $R\rightarrow 0$, and thus the BR factors $\bar{A}^{\rm(I,II)}_{xy}$
and $\bar{B}^{\rm(I,II)}_{xy}$ with spherical space regions vanish 
when $R=0$. 
The Fourier--Bessel expansions of the BR geometric factors, 
except those of the geometric factor $A^{\rm(I,II)}_{xx}$ for spherical
space regions whose centres coincide,
converge rapidly as less than a hundred terms were required for the 
four-digit accuracy with which the geometric factors are printed in table 1.

The slow convergence of expansion (67) in the case of spherical
space regions with coinciding centres
is due to the presence of a $q$-independent term in the quantity 
$\bar{\tilde{U}}_{00}^{(A_{xx})}(q)$ of equation (68). 
However, the contribution of this term 
to the BR geometric factor $\bar{A}^{\rm(I,II)}_{xx}$ has then a simple
form and can be summed easily.
With sphere separation $R=0$, it follows from equations (67) and (68)  
that the $q$-independent term contributes to the BR factor 
$\bar{A}^{\rm(I,II)}_{xx}$ the quantity
\begin{equation}
\bar{A}^{\rm(I,II)}_{xx}(g_0^{(0)}) =\frac{12g_0^{(0)}}{\pi {R_1}'R_2}
\frac{\pi}{r_{\rm ex}}\sum_{n=1}^{\infty}
j_1\left(\frac{n\pi}{r_{\rm ex}}{R_1}'\right)
j_1\left(\frac{n\pi}{r_{\rm ex}}R_2\right),
\end{equation}
where $g^{(0)}_0$, defined by equation (90), is the $q$-independent
term in question. 
The series in (98) is a Fourier--Bessel representation of 
of the simplest of the integrals evaluated in section 3.2:
\begin{eqnarray}
\frac{\pi}{r_{\rm ex}}\sum_{n=1}^{\infty}
j_1\left(\frac{n\pi}{r_{\rm ex}}{R_1}'\right)
j_1\left(\frac{n\pi}{r_{\rm ex}}R_2\right)&=&\int_0^{\infty}j_1({R_1}'q)
j_1(R_2q)\,dq\nonumber \\ 
&=&{\rm ji}_4(0;1,1,0,0;R'_1,R_2,0,0)=\frac{R_<}{R_>^2}\frac{\pi}{6}.
\end{eqnarray}
Here, the result (82) is simplified using $R_<$ and $R_>$, which are 
respectively the lesser and the greater 
of the radii ${R_1}'$ and $R_2$, and the parameter $r_{\rm ex}$ should be
such that $r_{\rm ex}\ge R_>$, which condition is guaranteed
by that of equation (67b).
When the result (99) is used in the expansion (67) in the cases of sphere 
separation $R=0$, the rate of convergence improves dramatically and
becomes similar to that of the geometric factors for space regions with 
noncoinciding centres.

The accuracy of the numerical results of table 1, obtained using the 
Fourier--Bessel expansions, was checked using the closed-form  
expressions (95)--(97). 
As no calculated values of the BR geometric factors could be found in the 
literature, this is the only check of the correctness and accuracy of our 
results. Admittedly, this check is not a fully independent one, as both
the closed-form expressions and Fourier--Bessel expansions 
are obtained using Fourier-transform methods. However, the calculations 
reported here use only proven analytical methods, of which  
Fourier integrals and Fourier--Bessel expansions are a part, 
and thus our results would be invalidated only if the same algebraic 
errors were made in the derivation of the closed-form and Fourier--Bessel 
expressions. In this connection, we note that the  
analytical expressions (48) and (49) for the double time averages, used  
in both the Fourier-integral and Fourier--Bessel formulations, were
checked by doing the two-dimensional integrations involved numerically.

A rather interesting result of the calculations is that the geometric factors
$\bar{A}^{\rm(I,I)}_{xx}$ with fully coinciding space-time regions turn out 
to have negative values. The closed-form expression (95) for 
the geometric factor $\bar{A}^{\rm(I,II)}_{xx}$ with radii $R_1=R_2=R_0$, 
time intervals $\Delta t_1=\Delta t_2=\Delta t_0$ and separations $R=T=0$ 
simplifies to\footnote{Using this simple expression, it can be shown that,
contrary to a conclusion of \cite{CP}, the BR average `self-force' on
the field measurement's test body approximates correctly the average self-force
that obtains when the duration of the momentum measurements on a test body
of sufficiently great mass is sufficiently short \cite{Com}.}
\begin{equation}
\bar{A}^{\rm(I,I)}_{xx}=-\frac{1}{8R_0^4\kappa}
(4+\kappa)(2-\kappa)^2\Theta(2-\kappa)-\frac{1}{R_0^4\kappa},
\end{equation}
where $\kappa=\Delta t_0/R_0$. 
For a fixed value of $R_0$, this function of the ratio $\kappa$ increases 
monotonically from $-\infty$ when $\kappa\rightarrow 0$ to the value 
of zero for $\kappa\rightarrow 0$. For $\kappa\ge 2$, the geometric factor
$\bar{A}^{\rm(I,I)}_{xx}$ reduces to the value $-1/R_0^4\kappa$,
and so the BR average `self-force' \cite{BR} on the field-measurement's
test body of charge density $\rho_{\rm I}$ is then
$\rho_{\rm I}^2 V_{\rm I}^2 \Delta t_0 D^{\rm (I)}_x\bar{A}^{\rm(I,I)}_{xx}
=-\rho_{\rm I}^2 V_{\rm I}^2 D^{\rm (I)}_x/R^3_0$, which is simply
the electrostatic force of attraction between the test and neutralizing bodies
when their centres are displaced by a distance $|D^{\rm(I)}_x|\ll R_0$.
The negativity of the BR geometric factor $\bar{A}^{\rm(I,I)}_{xx}$ 
means that the spring constant
$k_{\rm I}=\rho^2_{\rm I}V^2_{\rm I}T_{\rm I}\bar{A}^{\rm(I,I)}_{xx}$ 
of the spring that is used in a BR 
field measurement involving a space-time region $V_{\rm I},T_{\rm I}$ 
to compensate the test body's average `self-force' has to be negative.   
While it is certainly possible to envisage a spring mechanism that
would provide a force proportional to and in the direction of a test body's
displacement,\footnote{ 
A spring system with a negative spring constant can be constructed
as follows. Two elastically compressible rods, each of spring constant $k$ 
and length $l+d$ when not stressed, are aligned ``head to tail" along 
a common axis and joined via a movable joint, while their outer ends are 
fastened using similar joints to rigid supports so that this spring system 
is compressed to a total length $2l$. It can be shown easily that 
a body attached to the middle joint and moved a small distance 
$x$, $x\ll l$, $x\ll d$, from the system's axis and in a direction 
perpendicular 
to it, will experience a force $F=(2kd/l)x=-\kappa x$, which is 
proportional to and acting in the direction of the displacement $x$,
i.e., the spring constant $\kappa=-2kd/l$ of such a system is negative.}
we note that Bohr and Rosenfeld did not
consider it necessary to make a comment on this rather unusual specification
that their measurement procedure would place on the spring mechanism---but
one can now only speculate whether Bohr and Rosenfeld were in fact aware
of this consequence of their analysis.\footnote{
A hint that Bohr and Rosenfeld were aware of the possibility 
of the geometric factor $\bar{A}^{\rm(I,I)}_{xx}$ being negative is given
by their careful writing of its square root as 
$|\bar{A}^{\rm(I,I)}_{xx}|^{1/2}$.}
In any case, despite its inherent instability, a spring mechanism with  
negative spring constant should present no difficulty of principle for
a BR measurement procedure because a BR spring, together with the test body 
to which it is attached, is supposed to be released only 
for the exact duration of the field measurement, 
and the spring force is designed so that its effect is compensated
by the test body's `self-force'.

We conclude that a well-controlled method for the computation
of the BR geometric factors was developed using Fourier--Bessel expansions. 
The efficiency and accuracy of the method were tested numerically in the case 
of spherical space regions when a BR geometric factor
can be represented by a Fourier--Bessel series with 
terms expressed entirely in terms of elementary functions, and, 
using the computing system {\it Mathematica}, it is possible to 
obtain the factor in terms of manageable closed-form expressions.                         
The space-time-averaged electromagnetic-field commutators, as well 
as the formal expressions and {\it ``gedanken"} experimental procedures 
of the famous
Bohr--Rosenfeld analysis of the measurability of the electromagnetic field
now can be translated easily and accurately into concrete numbers. 

\begin{flushleft} 
\bf Acknowledgement
\end{flushleft}
The author is grateful to F. Frescura for a careful reading of the
manuscript. He also acknowledges gratefully correspondence with F. Persico,
whose searching questions helped the author to 
realize that the formal expressions for some BR geometric factors
imply a suitable regularization.
\vspace{20ex}

\begin{table}
\caption{Representative BR geometric factors $\bar{C}^{\rm(I,II)}_U$
for space-time regions I and II specified  
by space spheres of radii $R_1$ and $R_2$, and time intervals $\Delta t_1$ 
and  $\Delta t_2$, respectively, with the centre of the second sphere 
displaced from that of the first one by a vector of spherical 
coordinates $R$, $\theta$, $\phi$, and the beginning of the second time 
interval separated from that of the first one by a time interval $T$;
units such that the speed of light $c=1$ are used.} 
\begin{tabular}{ccccccccrr}
  & $R_1$ & $R_2$ & $R$ & $\theta$ & $\phi$ & 
$\Delta t_1$ & $\Delta t_2$ & $T$ & $\bar{C}^{\rm(I,II)}_U$  \\
\tableline
$\bar{A}^{\rm(I,II)}_{xx}$ &1 &1 &0 &-- &-- &1 &1 &0 &
$-1.625\times 10^{+0}\,\,\,\,$ \\
$\bar{A}^{\rm(I,II)}_{xx}$ &10 &10 &0 &-- &-- &1 &1 &0 &
$-2.850\times 10^{-3}\,\,\,\,$ \\
$\bar{A}^{\rm(I,II)}_{xx}$ &1 &1 &0 &-- &-- &1 &2 &0.5 &
$1.953\times 10^{-1}\,\,\,\,$ \\
$\bar{A}^{\rm(I,II)}_{xx}$\tablenotemark[1] &1 &1 &0 &-- &-- &2 &1 &$-0.5$ &
$-5.664\times 10^{-1}\,\,\,\,$\\
$\bar{A}^{\rm(I,II)}_{xx}$ &1 &1 &1 &$\case{1}{6}\pi$ &$\case{1}{3}\pi$ &1 
&1 &0.5 & $-6.407\times 10^{-2}\,\,\,\,$ \\
$\bar{A}^{\rm(I,II)}_{xx}$\tablenotemark[1] &1 &1 &1 &
$\case{5}{6}\pi$ & $\case{4}{3}\pi$ &1 
&1 &$-0.5$ & $-4.530\times 10^{-1}\,\,\,\, $ \\
$\bar{A}^{\rm(I,II)}_{xy}$ &1 & 1&1 &$\case{1}{6}\pi$ &$\case{1}{3}\pi$ & 1
&1 &0.5 & $6.636\times 10^{-2}\,\,\,\,$  \\
$\bar{A}^{\rm(I,II)}_{xy}$\tablenotemark[1] &1 & 1&1 &$\case{5}{6}\pi$ & 
$\case{4}{3}\pi$ & 
1 &1 &$-0.5$ & $5.901\times 10^{-3}\,\,\,\,$  \\
$\bar{B}^{\rm(I,II)}_{xy}$ &1 & 1&1 &$\case{1}{6}\pi$ &$\case{1}{3}\pi$ & 1
&1 &0.5 & $-2.730\times 10^{-1}\,\,\,\,$  \\
$\bar{B}^{\rm(I,II)}_{xy}$\tablenotemark[1] &1 & 1&1 &$\case{5}{6}\pi$ & 
$\case{4}{3}\pi$ & 
1 &1 &$-0.5$ & $1.675\times 10^{-1}\,\,\,\,$  \\
$\bar{A}^{\rm(I,II)}_{xx}$ &1 &2&1 &$\case{1}{6}\pi$ &$\case{1}{3}\pi$ &1 
&2 &0.5 & $7.454\times 10^{-2}\,\,\,\,$ \\
$\bar{A}^{\rm(I,II)}_{xx}$\tablenotemark[1] &2 &1 &1 &$\case{5}{6}\pi$ &
$\case{4}{3}\pi$ &2 &1 &$-0.5$ & $-8.914\times 10^{-2}\,\,\,\,$ \\
$\bar{A}^{\rm(I,II)}_{xy}$ &1 &2 &1 &$\case{1}{6}\pi$ &$\case{1}{3}\pi$ &1 
&2 &0.5 & $3.493\times 10^{-3}\,\,\,\,$  \\
$\bar{A}^{\rm(I,II)}_{xy}$\tablenotemark[1] &2 &1 &1 &$\case{5}{6}\pi$ &
$\case{4}{3}\pi$ &2 &1 &$-0.5$ & $-3.884\times 10^{-4}\,\,\,$ \\
$\bar{B}^{\rm(I,II)}_{xy}$ &1 & 2&1 &$\case{1}{6}\pi$ &$\case{1}{3}\pi$ & 1
&2 &0.5 & $-2.560\times 10^{-2}\,\,\,\,$  \\
$\bar{B}^{\rm(I,II)}_{xy}$\tablenotemark[1] &2 & 1&1 &$\case{5}{6}\pi$ & 
$\case{4}{3}\pi$ & 
2 &1 &$-0.5$ & $4.126\times 10^{-3}\,\,\,\,$ 
\end{tabular}
\tablenotetext[1]{The ``reverse" geometric factor $\bar{C}^{\rm(II,I)}_U$ 
                  of the preceding entry.}
         
\end{table}

\nopagebreak[4]


\begin{references}
\bibitem{BR} 
Bohr N and Rosenfeld L 1933 Zur Frage der Messbarkeit der
elektromagnetischen Feldgr{\"o}ssen {\it Mat.-fys. Medd. Dan. Vid. Selsk.}
{\bf 12} no 8; Engl. transl. 1979 On the question of the
measurability of electromagnetic field quantities 
{\it Selected Papers of L{\'e}on Rosenfeld}  eds R S Cohen and J Stachel 
(Dordrecht: Reidel) pp 357--400, this translation reprinted 
1983  {\it Quantum Theory and Measurement} eds J A Wheeler and W H Zurek
(Princeton, New Jersey: Princeton U.P.) pp 479--522
\bibitem{AP}
Pais A 1991 {\it Niels Bohr's Times, In Physics, Philosophy, and Polity} 
(Oxford: Clarendon Press) ch 16 (d)
\bibitem{LP}
Landau L and Peierls L 1931 Erweiterung des Unbestimmtheitsprinzips
f{\"u}r die relativistische Quantentheorie {\it Z. Phys.} {\bf 69}
56--69; Engl. transl. 1965 Extension of the uncertainty principle to
relativistic quantum theory {\it Collected Papers of
L. D. Landau} ed D Ter Haar (Oxford: Pergamon Press) pp 40--51; 
this transl. reprinted 
1983 {\it Quantum Theory and Measurement} eds J A Wheeler and W H Zurek 
(Princeton, New Jersey: Princeton U.P.) pp 465--476
\bibitem{AB}
Aharonov Y and Bohm D 1961 Time in quantum theory and the uncertainty
relation for time and energy {\it Phys. Rev.} {\bf 122} 1649--58,
reprinted 
1983 {\it Quantum Theory and Measurement} eds J A Wheeler and W H Zurek 
(Princeton, New Jersey: Princeton U.P.) pp 715--724\\
Aharonov Y and Petersen A 1971 Definability and measurability in
quantum theory {\it Quantum Theory and Beyond} ed T Bastin  
(Cambridge: Cambridge U.P.) pp 135--139
\bibitem{Rosen} 
Rosenfeld L 1965 On quantum electrodynamics   
{\it Niels Bohr and the Development of Physics} ed W Pauli
(London: Pergamon Press) pp 70--95
\bibitem{Peierls}  
Peierls R 1985 {\it Bird of Passage: Recollections of a Physicist}
(Princeton, New Jersey: Princeton U.P.) pp 66--67 
\bibitem{Miller} 
Miller A I 1990 Measurement problems in quantum field theory in the
1930's {\it Sixty-Two Years of Uncertainty} ed A I Miller 
(New York: Plenum Press) pp 139--152
\bibitem{CP}
Compagno G and Persico F 1998 Limits on the measurability of the local
quantum electromagnetic-field amplitude {\it Phys. Rev. A} {\bf 57}
1595--1603
\bibitem{Com} Hnizdo V 1999 Comment on ``Limits of the measurability of the 
local quantum electromagnetic-field amplitude" {\it Phys. Rev. A} {\bf 60}
4191--4195; quant-ph/0210074
\bibitem{VH} 
Hnizdo V 1994 Evaluation of folding integrals using Fourier--Bessel
expansions  {\it J. Phys. A} {\bf 27} 7139--7145
\bibitem{QNM}
Braginsky V B, Vorontsov Y I and Thorne K S 1980
Quantum nondemolition measurements {\it Science} {\bf 209} 547--557,
reprinted 
1983  {\it Quantum Theory and Measurement} eds J A Wheeler and W H Zurek
(Princeton, New Jersey: Princeton U.P.) pp 749--768\\
Braginsky V B and Khalili F Y 1996 Quantum nondemolition experiments:
The route from toys to tools {\it Rev. Mod. Phys.} {\bf 68} 1--11\\
Bocko M F and Onofrio R 1996 On the measurement of a weak classical
force coupled to a harmonic oscillator: Experimental progress
{\it Rev. Mod. Phys.} {\bf 68} 755--799
\bibitem{Arf} Arfken G 1970 {\it Mathematical Methods for Physicists} 
2nd edn (New York: Academic) p 681
\bibitem{Wolfram}{\it Mathematica} 3.0 1996 (Champaign, Il.: 
Wolfram Research, Inc.) 
\end{references}
\end{document}